\PassOptionsToPackage{svgnames, table}{xcolor}

\documentclass[12pt, letterpaper]{article}

\usepackage[english]{babel}
\usepackage{amsmath,amssymb,amsthm}
\usepackage[redefine-symbols=false]{siunitx}
\usepackage{xfrac} 
\usepackage{caption}
\pdfpageattr {/Group << /S /Transparency /I true /CS /DeviceRGB>>}
\usepackage{xcolor}
\usepackage{graphicx}
\usepackage{booktabs}
\usepackage{fancyhdr}  
\usepackage[compact]{titlesec}  
\usepackage[
unicode,
pdftex,
bookmarks=true,
bookmarksnumbered=true,
bookmarksopen=true,
bookmarksopenlevel=2,
colorlinks,
linkcolor=blue!80!black!80,
citecolor=blue!40!black!90,
linktocpage=false,
urlcolor=cyan,
]{hyperref}

\let\orgautoref\autoref

\renewcommand{\autoref}
        {\def\equationautorefname{Eq.}%
         \def\figureautorefname{Fig.}%
         \def\subfigureautorefname{Fig.}%
         \def\sectionautorefname{\S}
         \def\subsectionautorefname{\S}%
         \def\subsubsectionautorefname{\S}%
         \def\Itemautorefname{item}%
         \def\tableautorefname{Table}%
         \orgautoref}

\usepackage[%
open,
openlevel=2
]{bookmark}[2011/04/21]

\usepackage{apacite}
\usepackage{braket}

\DeclareMathOperator\var{Var}
\DeclareDocumentCommand{\atime}{m m}{E[T_{\mathtt{#1} | \mathtt{#2}}]}

\renewcommand{\geq}{\geqslant}
\newcommand{\HH}{\mathtt{HH}}
\newcommand{\HT}{\mathtt{HT}}


\titleformat{name=\section}
{\filcenter\Large\bfseries}
{}
{.5em}
{}
\titleformat{name=\section,numberless}
{\filcenter\Large\bfseries}
{}
{.5em}
{}

\let\chapter\undefined

\renewcommand{\sectionmark}[1]%
{\markright{\MakeUppercase{#1}}}

\newcommand{\helv}{%
\fontfamily{phv}\fontseries{b}\fontsize{9}{11}\selectfont}
\makeatletter
\renewcommand\@seccntformat[1]{}
\makeatother

\begin{document}

\pagenumbering{roman}
\setcounter{page}{1}

\begin{titlepage}
	\centering
	\vspace{1.5cm}
    {\Large\bfseries Learning Temporal Structures of Random Patterns \par}	

	\vspace{0.8cm}
    {Yanlong Sun \& Hongbin Wang \par}
    \vspace{0.2cm}
	{College of Medicine, Texas A\&M University \par}
    \vspace{0.5cm}

    \vspace{1.5cm}
    {\large\bfseries Abstract \par}
    \vspace{0.3cm}
    \begin{minipage}{0.9\linewidth}
        A cornerstone of human statistical learning is the ability to extract temporal regularities / patterns from random sequences.
        Here we present a method of computing pattern time statistics with generating functions for first-order Markov trials and independent Bernoulli trials.
        We show that the pattern time statistics cover a wide range of measurements commonly used in existing studies of both human and machine learning of stochastic processes, including probability of alternation, temporal correlation between pattern events, and related variance / risk measures.
        Moreover, we show that recurrent processing and event segmentation by pattern overlap may provide a coherent explanation for the sensitivity of the human brain to the rich statistics and the latent structures in the learning environment.
    \begin{flushleft}
    \end{flushleft}
    \end{minipage}

\end{titlepage}

\pagenumbering{arabic}
\setcounter{page}{1}

\fancyhead[L]{\helv \rightmark}
\fancyhead[R]{\helv \thepage}
\renewcommand{\subsectionmark}[1]{}

\section{Introduction}
Random events are ubiquitous in both everyday life and scientific endeavors.
Reasoning about randomness, however, can be tricky and often counter-intuitive.
To illustrate, consider a game where two players, Alice and Bob, each have a fair coin and begin to flip it repeatedly.
Alice is waiting for pattern $\mathtt{HT}$ (a head followed by a tail) and Bob is waiting for pattern $\mathtt{HH}$ (a head followed by a head).
Suppose that on the first flip they both get an $\mathtt{H}$.
Then who is more likely to \emph{first} obtain her or his desired pattern with fewer flips?

Since the outcome of the second flip can be either a $\mathtt{T}$ or an $\mathtt{H}$ with equal probability, $p(\mathtt{T|H}) = p(\mathtt{H|H}) = \sfrac{1}{2}$, one might expect a tie.
The correct answer, however, is that Alice is more likely to win the contest with on average two fewer flips than Bob.
Let $E[T_{j|i}]$ denote the expected number of flips until the first arrival of pattern $j$ starting with pattern $i$, we have $\atime{HT}{H}=2$ and $\atime{HH}{H}=4$.
Interestingly, the expectation of a tie may still hold---if the game is played with a single coin where two players are waiting for their respective patterns from the same sequence of flips.

Examples of this sort abound, highlighting the fact that standard probability theory can sometimes fail to capture a richer body of latent structures embedded in random sequences.
It is remarkable that there is evidence that the human brain is capable of capturing some of these structures and demonstrate its effects in behavior, which, ironically, are often called cognitive biases. Examples include the representativeness heuristic, the law of small numbers, the gambler’s fallacy and hot hand belief \cite<e.g.,>{Budescu1987,Gilovich1985,Rabin2010,Tversky1971,Tversky1974,Wagenaar1972}.
However, by pattern time statistics such as $\atime{HT}{H} < \atime{HH}{H}$, this particular ``bias'', where one assumes that given an initial $\mathtt{H}$, a $\mathtt{T}$ is more imminently due than $\mathtt{H}$,  may be a consequence of temporal structure learning in the brain --- the initial $\mathtt{H}$ becomes more strongly associated with $\mathtt{HT}$ than $\mathtt{HH}$ because $\mathtt{HH}$ is ``delayed'' \cite{Sun2015pnas,Sun2010cogpsy,Sun2015CogSci,sun2017cogsci}.
It is important that we pursue how temporal structures embedded in random patterns can be learned, by both the human brain and machines.

In this paper, we present a systematic treatment of pattern time statistics by the method of generating functions.
Our motivation is mainly twofold.
First, we would like to provide an intuitive and streamlined method for computing pattern time statistics, then use the result to give a coherent interpretation to other statistical measures in existing studies on human perception of randomness.
Second, computing pattern time statistics by generating functions is in effect a theory of statistical learning that can be applied in both cognitive and artificial systems.
It has been suggested that subjective probability estimates are sensitive to the spatio-temporal distances \cite<e.g.,>{Luhmann2008,McClure2004,Trope2010}.
The way a generating function organizes combinatorial objects then compresses the representation to produce a certain statistic sheds new insights on newly proposed learning mechanisms in cognitive neuroscience and AI \cite{Elman1990,Marr1982,LeCun2015-NatureDeepReview,OReilly2014TI,Tenenbaum2011}.

\section{Pattern Time Statistics by Generating Functions}

Our method of generating functions for computing pattern time statistics is based on \citeA{Graham1994}, generalized from independent Bernoulli trials to first-order dependent Markov trials with a flexible event segmentation by auxiliary states.
Before proceeding, we should note that there is a variety of terminology in literature \cite<e.g.,>{Chang2005,Feller1968,Fu2002,Gardner1988,Li1980,Nickerson2007,Ross2007}.
Here we define a pattern time in its most general form, $T_{j|i} \geq 1$, as the random variable denoting the number of transitions for a random process to travel from the initial pattern $i$ until the \emph{first} arrival at the destination pattern $j$.
Two special cases are given unique names: $T_{j|\varnothing}$ is the \emph{first-arrival time} of pattern $j$ from an empty initial state $\varnothing$ (i.e., the process starts anew), and its expected value $E[T_{j|\varnothing}]$ is the pattern's \emph{waiting time};
$T_{j|j}$ is the \emph{inter-arrival time} between any two consecutive occurrences of pattern $j$, and its expected value $E[T_{j|j}]$ is the pattern's \emph{mean time}.

\subsection{Generating Functions}

Following the notation by \citeA{Graham1994}, a \emph{generating function}, $A(z)$, is the sum of a power series that organizes an infinite sequence $\{ a_0, a_1, a_2, \ldots \}$ with an auxiliary variable $z$,
\begin{equation}\label{eq:Az-power-series}
  A(z) = a_0 + a_1 z + a_2 z^2 + \cdots = \sum_{k \geq 0} a_k z^k .
\end{equation}
Then, a \emph{probability generating function}, $G_X(z)$, where $X$ is a random variable that takes only nonnegative integer values, is the sum of the probability distribution,
\begin{equation}\label{eq:graham-pgf-def}
  G_X (z) = \sum_{k\geq 0} \mathrm{Pr}(X=k) z^k \, ,
\end{equation}
where $G_X (1) = \sum_{k\geq 0} \mathrm{Pr}(X=k) = 1$ represents the constraint that the total probability sums to one.
The power series in $G_X (z)$ contains all the information about the distribution of $X$.
Here we are only interested in the mean and variance, which are given by the first and second derivatives of $G_X(1)$,
\begin{align}\label{eq:pgf-and-E-V}
  \begin{split}
    E[X] &= G_X'(1), \\
    \var(X) &=  G_X''(1) + G'_X(1) - G'_X(1)^2.
  \end{split}
\end{align}
Then, to compute $E[T_{j|i}]$ and $\var(T_{j|i})$, we simply need to find the corresponding probability generating function for the random variable $T_{j|i}$.

\subsection{First-order Markov Trials}
First-order dependent Markov trials has been a widely used model in studies on human randomness perception \cite<e.g.,>{Budescu1987,Falk1997,Lopes1987,Nickerson2002,Oskarsson2009,Sun2012cogsci,Sun2015pnas}.
Assume that the process is $\mathtt{H}$-$\mathtt{T}$ symmetrical with stationary probabilities,
\begin{equation*}
  \pi_{\mathtt{H}} = \pi_{\mathtt{T}} = \sfrac{1}{2}.
\end{equation*}
This means that the first transition out of an empty initial state $\varnothing$ has an equal chance to end up with an $\mathtt{H}$ or $\mathtt{T}$.
Let $p_A$ denote the \emph{probability of alternation} between consecutive trials, then all the transition probabilities can be simplified as,
\begin{equation*}
  p_A = p(\mathtt{T|H}) = p(\mathtt{H|T}) = 1- p(\mathtt{H|H}) = 1- p(\mathtt{T|T}).
\end{equation*}

\begin{figure}[htb]\centering
\includegraphics[scale=0.9]{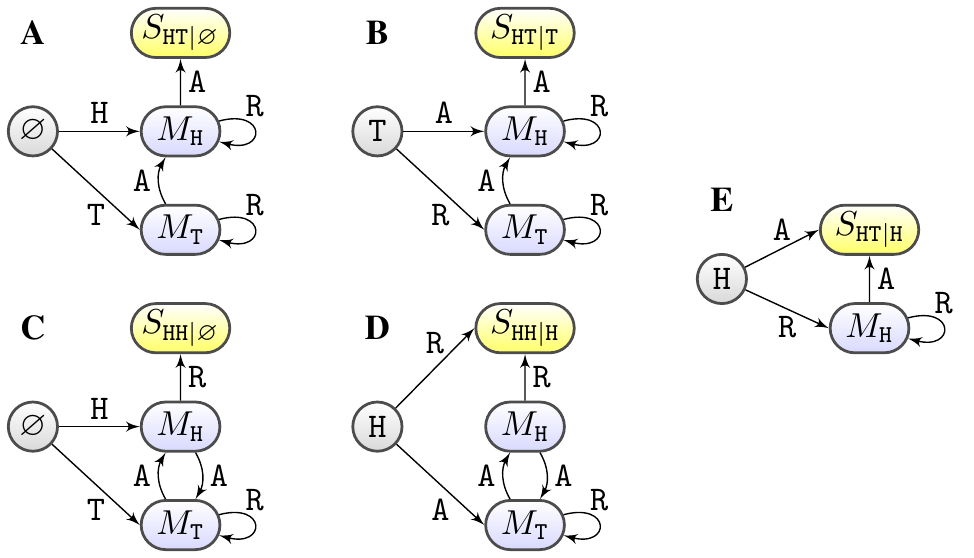}
\caption{Markov chains for generating the distributions of $T_{j|i}$ by first-order Markov trials.
In each chain, destination states $S_{j|i}$ represent all possible sequences that start from pattern $i$ and end with the first arrival of pattern $j$, and auxiliary states $M_k$ represent all possible sequences that end with pattern $k$ but do not contain the destination pattern $j$.
Transitions between nonempty states are labeled as either repetition ($\mathtt{R}$) or alternation ($\mathtt{A}$).}
\label{fig:MarkovChains}
\end{figure}

Our first example is to derive the distribution of $T_{\mathtt{HT|\varnothing}}$.
Let $S_{\mathtt{HT|\varnothing}}$ denote the sum of all sequences that end with the first arrival of the destination pattern $\mathtt{HT}$, $M_{\mathtt{T}}$ denote the sum of all sequences that end with a $\mathtt{T}$ but do not contain any $\mathtt{HT}$, and $M_{\mathtt{H}}$ denote the sum of all sequences that end with an $\mathtt{H}$ but do not contain any $\mathtt{HT}$.
Each of $S_{\mathtt{HT|\varnothing}}$, $M_{\mathtt{T}}$ and $M_{\mathtt{H}}$ is a generating function that can be represented as a state in a Markov chain, where $S_{\mathtt{HT|\varnothing}}$ is the destination state, and $M_{\mathtt{T}}$ and $M_{\mathtt{H}}$ are two auxiliary states (\autoref{fig:MarkovChains}A).
To see the exact composition of a generating function, take $M_{\mathtt{T}}$ as an example,
\begin{equation}\label{eq:Markov-GF-0-HT-Mt}
  M_{\mathtt{T}} = \mathtt{T} + \mathtt{TR} + \mathtt{TRR} + \mathtt{TRRR} + \cdots
  = \mathtt{T} \sum_{n \geq 0} \mathtt{R}^n = \frac{\mathtt{T}}{1-\mathtt{R}} ,
\end{equation}
where $\mathtt{R}$ represents a transition of repetition (e.g., $\mathtt{TR}$ results in the sequence $\mathtt{TT}$).
This shows that $M_{\mathtt{T}}$ is the sum of a power series so that a closed form of $M_{\mathtt{T}}$ can be obtained by the geometric distribution $\sum_{n \geq 0} z^n = \frac{1}{1-z}$.
Equivalently, $M_{\mathtt{T}}$ can be decomposed in a recursive structure with respect to itself,
\begin{equation}\label{eq:Markov-GF-0-HT-Mt-rel}
  M_{\mathtt{T}} = \mathtt{T} + (\mathtt{T} + \mathtt{TR} + \mathtt{TRR} + \cdots ) \mathtt{R}
  = \mathtt{T} + M_{\mathtt{T}} \mathtt{R} \, .
\end{equation}
Solving $M_{\mathtt{T}} = \mathtt{T} + M_{\mathtt{T}} \mathtt{R}$ gives us the same result as \autoref{eq:Markov-GF-0-HT-Mt}.
The equivalence between \autoref{eq:Markov-GF-0-HT-Mt} and \autoref{eq:Markov-GF-0-HT-Mt-rel} is a consequence of the memoryless property of Markov trials, by which the transition out of a state is independent of the transition into that state.
As such, a generating function follows a rule of ``multiplication by juxtaposition'', for example, event $\mathtt{TR}$ is the product of two independent events $\mathtt{T}$ and $\mathtt{R}$.
Note that by the arrow of time, $\mathtt{TR} \neq \mathtt{RT}$.

We can directly construct $S_{\mathtt{HT}|\varnothing}$ in a way similar to \autoref{eq:Markov-GF-0-HT-Mt} or \autoref{eq:Markov-GF-0-HT-Mt-rel}, but it would be easier to follow its Markov chain in \autoref{fig:MarkovChains}A and write
\begin{align}\label{eq:Markov-GF-0-to-HT-rel}
  \begin{split}
    M_{\mathtt{T}} &= \mathtt{T} + M_{\mathtt{T}} \mathtt{R} \, ,\\
    M_{\mathtt{H}} &= \mathtt{H} + M_{\mathtt{H}} \mathtt{R} + M_{\mathtt{T}} \mathtt{A} \, ,\\
    S_{\mathtt{HT}|\varnothing} &= M_{\mathtt{H}} \mathtt{A} \, ,
  \end{split}
\end{align}
where $\mathtt{A}$ represents an alternation and $\mathtt{R}$ represents a repetition.
Solve for $S_{\mathtt{HT}|\varnothing}$,
\begin{equation}\label{eq:Markov-GF-0-to-HT}
  S_{\mathtt{HT}|\varnothing} = \frac{\mathtt{HA}-\mathtt{HRA} + \mathtt{TAA}}{(1-\mathtt{R})^2}.
\end{equation}
Replace each $\mathtt{H}$ and each $\mathtt{T}$ with $\sfrac{z}{2}$ (since $\pi_{\mathtt{H}} = \pi_{\mathtt{T}} = \sfrac{1}{2}$), each $\mathtt{A}$ with $p_A z$, and each $\mathtt{R}$ with $(1-p_A)z$, we have the probability generating function
\begin{equation}\label{eq:Markov-PGF-0-to-HT}
  G_{\mathtt{HT}|\varnothing}(z) = \frac{p_A (2p_A z - z + 1) z^2}{2 (p_A z-z+1)^2}.
\end{equation}
By \autoref{eq:pgf-and-E-V}, we have
\begin{equation}\label{eq:Markov-EV-0-to-HT}
  E[T_{\mathtt{HT}|\varnothing}] = 1 + \frac{1}{2p_A} + \frac{1}{p_A} , \quad
  \var(T_{\mathtt{HT}|\varnothing}) = \frac{7}{4 p_A^2} - \frac{3}{2 p_A} . 
\end{equation}
For example, at $p_A = \sfrac{1}{2}$, we have $E[T_{\mathtt{HT}|\varnothing}] = 4$ and $\var(T_{\mathtt{HT}|\varnothing}) = 4$.

To recapitulate, the procedure we just described can be summarized as three levels of aggregation, from which abstract representations are extracted from concrete combinatorial objects.
First, a generating function groups sequences of the same property as power series into a relational structure.
Then, a probability generating function discards the exact grouping information so that only the sequence length is preserved in the power series of $z$.
Finally, averaging all sequence lengths at $z=1$, we obtain $E[T_{\mathtt{HT}|\varnothing}]$ and $\var(T_{\mathtt{HT}|\varnothing})$.

Apply the same technique to the distribution of $T_{\mathtt{HH|\varnothing}}$ (\autoref{fig:MarkovChains}C), we have
\begin{align}\label{eq:Markov-GF-0-to-HH-eqs}
  \begin{split}
    M_{\mathtt{T}} &= \mathtt{T} + M_{\mathtt{T}} \mathtt{R} + M_{\mathtt{H}} \mathtt{A} \, , \\
    M_{\mathtt{H}} &= \mathtt{H} + M_{\mathtt{T}} \mathtt{A} \, , \\
    S_{\mathtt{HH}|\varnothing} &= M_{\mathtt{H}} \mathtt{R} \, .
  \end{split}
\end{align}
Note that here $M_{\mathtt{T}}$ and $M_{\mathtt{H}}$ respectively denote the sums of sequences that end with $\mathtt{T}$ and $\mathtt{H}$ but do not contain $\mathtt{HH}$.
In general, an auxiliary state toward different destination states may have different compositions (cf., \autoref{eq:Markov-GF-0-to-HT-rel}).
Solve for $S_{\mathtt{HH}|\varnothing}$ then discover $G_{\mathtt{HH}|\varnothing}(z)$, we have
\begin{equation}\label{eq:Markov-EV-0-to-HH}
  E[T_{\mathtt{HH}|\varnothing}] = 1 + \frac{1}{2 p_A} + \frac{2}{1-p_A} , \quad
  \var(T_{\mathtt{HH}|\varnothing}) = \frac{3}{4 p_A^2} + \frac{4}{(1-p_A)^2} + \frac{3}{2 p_A} .
\end{equation}

At $p_A = \sfrac{1}{2}$, we have $E[T_{\mathtt{HH}|\varnothing}] = 6$ and $\var(T_{\mathtt{HH}|\varnothing}) = 22$.
Compared with $T_{\mathtt{HT}|\varnothing}$ from \autoref{eq:Markov-EV-0-to-HT}, this shows that at $p_A = \sfrac{1}{2}$, $T_{\mathtt{HH}|\varnothing}$ not only have a greater mean, but also a greater variance.
Moreover, $E[T_{\mathtt{HH}|\varnothing}] = E[T_{\mathtt{HT}|\varnothing}] = 5.5$ at $p_A = \sfrac{1}{3}$.
That is, alternations have to be half as frequent as repetitions to make patterns $\mathtt{HT}$ and $\mathtt{HH}$ have the same waiting time.
As shown in \autoref{fig:MarkovChains} (A and C), the final transition toward $S_{\mathtt{HH}|\varnothing}$ requires a repetition of the last $\mathtt{H}$ of a member in $M_{\mathtt{H}}$, but an alternation would set the process back into the loop between $M_{\mathtt{H}}$ and $M_{\mathtt{T}}$.
In contrast, transitions toward $S_{\mathtt{HT}|\varnothing}$ do not have such a delay.
This provides an intuitive explanation as to why $E[T_{\mathtt{HH}|\varnothing}] > E[T_{\mathtt{HT}|\varnothing}]$ at $p_A = \sfrac{1}{2}$.

Distributions of other pattern times in \autoref{fig:MarkovChains} can be obtained in the same way so here we only give the end results,
\begin{equation}\label{eq:Markov-EV-lump}
\begin{alignedat}{3}
  E[T_{\mathtt{HT|T}}] &= \frac{2}{p_A} , \quad  & \var(T_{\mathtt{HT|T}}) &= \frac{2}{p_A^2} - \frac{2}{p_A} ; \\[0.2em]
  E[T_{\mathtt{HH|H}}] &= \frac{2}{1 - p_A} , \quad  & \var(T_{\mathtt{HH|H}}) &= \frac{4}{(1-p_A)^2} + \frac{2}{p_A} ; \\[0.2em]
  E[T_{\mathtt{HT|H}}] &= \frac{1}{p_A} , \quad  & \var(T_{\mathtt{HT|H}}) &= \frac{1}{p_A^2} - \frac{1}{p_A}.
\end{alignedat}
\end{equation}

A notable comparison is between \autoref{fig:MarkovChains}B and D.
At $p_A = \sfrac{1}{2}$, we have $E[T_{\mathtt{HT|T}}] = E[T_{\mathtt{HH|H}}]  = 4$, $\var(T_{\mathtt{HT|T}}) = 4$ but $\var(T_{\mathtt{HH|H}}) = 20$, showing that the inter-arrival times of $\mathtt{HT}$ and $\mathtt{HH}$ have the same mean but different variances.
Note that the same mean time is equivalent to $p(\mathtt{T|H}) = p(\mathtt{H|H}) = \sfrac{1}{2}$ or $p(\mathtt{HT}) = p(\mathtt{HH}) = \sfrac{1}{4}$ by fair-coin Bernoulli trials, where the \emph{probability of occurrence} $p(\mathtt{HT})$ or $p(\mathtt{HH})$ is the inverse of the mean time, $p(\mathtt{HT}) = 1/ E[T_{\mathtt{HT|T}}]$ and $p(\mathtt{HH}) = 1/ E[T_{\mathtt{HH|H}}]$.

Another notable comparison is between \autoref{fig:MarkovChains}D and E.
At $p_A = \sfrac{1}{2}$, we have $E[T_{\mathtt{HT|H}}]  = 2$ and $E[T_{\mathtt{HH|H}}]  = 4$, which is the result of the game we discussed at the beginning of this paper.
This is because given the same initial state $\mathtt{H}$, the waiting for $\mathtt{HT}$ at any moment is to wait for a single alternation (note that $T_{\mathtt{HT|H}} = T_{\mathtt{T|H}}$), however, the waiting for $\mathtt{HH}$ can be delayed if the first transition does not turn out as desired.

\subsection{Independent Bernoulli Trials}

\begin{figure}[htb]\centering
\includegraphics[scale=1]{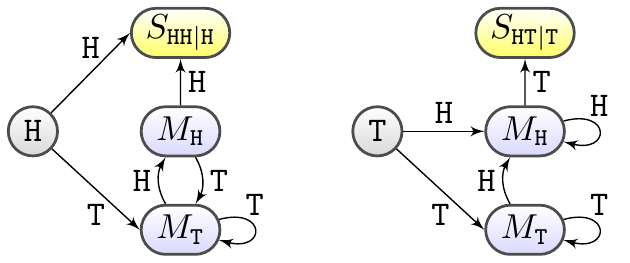}
\caption{Markov chains for generating the inter-arrival times $T_{j|j}$ for patterns $\mathtt{HH}$ and $\mathtt{HT}$ by independent Bernoulli trials. Destination states $S_{\mathtt{HH|H}}$ and $S_{\mathtt{HT|T}}$ represent all sequences that end with the first occurrence of the expected pattern, given that the pattern has just occurred.
Auxiliary states $M_{\mathtt{H}}$ and $M_{\mathtt{T}}$ represent all sequences that end with either an $\mathtt{H}$ and a $\mathtt{T}$, respectively, but do not contain the expected pattern.}
\label{fig:Bernoulli-GF-H-to-HH-and-T-to-HT}
\end{figure}

The method for independent Bernoulli trials is essentially the same as that for first-order Markov trials.
\autoref{fig:Bernoulli-GF-H-to-HH-and-T-to-HT} illustrates two examples of the Markov chains by Bernoulli trials, which can be obtained from the corresponding Markov chains in \autoref{fig:MarkovChains} by replacing $\mathtt{A}$ and $\mathtt{R}$ transitions with $\mathtt{H}$ and $\mathtt{T}$ transitions.
As a consequence of this replacement, the transformation from a generating function to its probability generating function is done by the substitution $\mathtt{H}=p_{\mathtt{H}} z$ and $\mathtt{T}=(1-p_{\mathtt{H}})z$, where $p_{\mathtt{H}}$ denotes the probability of heads.

The following are some of the end results by Bernoulli trials,
\begin{equation}\label{eq:Bernoulli-EV-lump}
\begin{alignedat}{3}
  E[T_{\mathtt{HT|\varnothing}}] &= \frac{1}{p_{\mathtt{H}}} + \frac{1}{1-p_{\mathtt{H}}} , \quad
  & \var(T_{\mathtt{HT|\varnothing}})
  &= \frac{1}{p_{\mathtt{H}}^2} - \frac{1}{p_{\mathtt{H}}} + \frac{1}{(1-p_{\mathtt{H}})^2} - \frac{1}{1-p_{\mathtt{H}}} ; \\[0.2em]
  E[T_{\mathtt{HH|\varnothing}}] &= \frac{1}{p_{\mathtt{H}}} + \frac{1}{p_{\mathtt{H}}^2} , \quad
  & \var(T_{\mathtt{HH|\varnothing}})
  &= \frac{1}{p_{\mathtt{H}}^4} + \frac{2}{p_{\mathtt{H}}^3} - \frac{2}{p_{\mathtt{H}}^2} -\frac{1}{p_{\mathtt{H}}} ; \\[0.2em]
  E[T_{\mathtt{HT|H}}] &= \frac{1}{1-p_{\mathtt{H}}} , \quad
  & \var(T_{\mathtt{HT|H}})
  &= \frac{1}{(1-p_{\mathtt{H}})^2} - \frac{1}{1-p_{\mathtt{H}}} ; \\[0.2em]
  E[T_{\mathtt{HH|H}}] &= \frac{1}{p_{\mathtt{H}}^2} , \quad
  & \var(T_{\mathtt{HH|H}})
  &= \frac{1}{p_{\mathtt{H}}^4} + \frac{2}{p_{\mathtt{H}}^3} - \frac{3}{p_{\mathtt{H}}^2} .
\end{alignedat}
\end{equation}

Note that $T_{\mathtt{HT|H}} = T_{\mathtt{T|\varnothing}}$ follows a geometric distribution, which means that given an initial $\mathtt{H}$, the waiting for pattern $\mathtt{HT}$ at any moment is only to wait for a single $\mathtt{T}$.
At $p_{\mathtt{H}} = \sfrac{1}{2}$, we have $E[T_{\mathtt{HT|H}}]  = 2$ and $E[T_{\mathtt{HH|H}}]  = 4$, which is the same result given by \autoref{eq:Markov-EV-lump}.
Moreover, $T_{\mathtt{HT|T}} = T_{\mathtt{HT|\varnothing}}$ means that the mean time of pattern $\mathtt{HT}$ is equal to its waiting time.
This is because given an occurrence of pattern $\mathtt{HT}$, its reoccurrence must start anew.

As an interesting extension of \autoref{eq:Bernoulli-EV-lump}, consider a game of roulette at the Monte Carlo casino in $1913$ when black repeated a record 26 times and people began extreme betting on red after about 15 repetitions \cite{HuffD1959}.
Given the same initial state of $k$ heads, let $T_{k\mathtt{H,H}|k\mathtt{H}}$ denote the first-arrival time until a streak of $k+1$ heads, and $T_{k\mathtt{H,T}|k\mathtt{H}}$ denote the first-arrival time until a streak of $k$ heads is terminated by an alternation at the end,
\begin{equation}\label{eq:AT-k-Bernoulli}
  E[T_{k\mathtt{H,H}|k\mathtt{H}}] = \frac{1}{p_{\mathtt{H}}^{k+1}},
  \quad  E[T_{k\mathtt{H,T}|k\mathtt{H}}] = E[T_{\mathtt{T}|\varnothing}] =  \frac{1}{1-p_{\mathtt{H}}}.
\end{equation}
This is because $E[T_{k\mathtt{H,H}|k\mathtt{H}}]$ is the mean time of a streak of $(k+1)$ heads, and $E[T_{k\mathtt{H,T}|k\mathtt{H}}]$ is merely the mean time or waiting time of a single tail.
When $p_{\mathtt{H}}=\sfrac{1}{2}$ and $k=15$, we have $E[T_{k\mathtt{H,H}|k\mathtt{H}}] = 2^{16}$ and $E[T_{k\mathtt{H,T}|k\mathtt{H}}]=2$.
However, this only means that the expected temporal distance $(k \mathtt{H} \to k \mathtt{H}, \mathtt{H})$ is greater than $(k \mathtt{H} \to k \mathtt{H}, \mathtt{T})$.
It does not mean that an existing streak is less likely to be extended by a repetition than to be terminated by an alternation, an expectation known as the gambler's fallacy \cite<e.g.,>{Tversky1971, Tversky1974}.

\section{Pattern Overlap, Event Segmentation and Sample Length}

The method of generating functions we just described extracts a pattern time statistic by aggregating over all possible sequences, including those of infinite length.
This raises a question whether it is physically or biologically plausible for a perceiving agent, with limited memory capacity, to actually learn such a statistic through limited exposure to the learning environment.
In the following, we show that the method of generating functions hinges on pattern overlap, which is an intrinsic property of the pattern.
For more complex patterns,  event segmentation by auxiliary states can partition the probability space with a limited number of recursive structures.
Consequently, the result of a generating function can be approximated by recursively applying the overlap property within sequences of finite length from which the pattern is sampled.

\subsection{Pattern Overlap and Event Segmentation}
For both first-order Markov trials and independent Bernoulli trials, with all other parameters fixed (e.g., pattern length, probability of alternation or probability of tossing heads), the distribution of $T_{j|i}$ is entirely determined by the \emph{overlap} between patterns $i$ and $j$, which is the maximal number of elements at the end of pattern $i$ that can be used as the beginning part of pattern $j$.
Based on this pattern overlap, the method of generating function we presented above can be easily generalized to any pattern of an arbitrary length with a proper \emph{event segmentation}.

To see this, consider the first-arrival time of an arbitrary binary pattern.
We only need two types of infinite sums.
Let $S$ denote the sum of all sequences that end with the pattern's first arrival, and $M$ denote the auxiliary sum of all non-empty sequences that contain no occurrence of the pattern.
Because any non-empty sequence belongs to either $S$ or $M$, and extending any member of $M$ with a single $\mathtt{H}$ or $\mathtt{T}$ results in a sequence in either $S$ or $M$, we have
\begin{equation}\label{eq:Bernoulli-aux-sum-M}
  S + M = M (\mathtt{H} + \mathtt{T}) + \mathtt{H} + \mathtt{T} \, .
\end{equation}
Suppose that the pattern of interest is $\underline{\mathtt{TH}} \mathtt{T} \underline{\mathtt{TH}}$, where the underlined elements $\mathtt{TH}$ are the overlap between any two immediate reoccurrences of the pattern.
This means that if any occurrence of $\mathtt{THTTH}$ is considered as a whole event, each event must begin and end with the same segment $\mathtt{TH}$.
Since any member of $S$ must end with $\mathtt{TH}$, appending $S$ with $\mathtt{TTH}$ results in sequences in which $\mathtt{THTTH}$ has occurred twice.
Therefore,
\begin{equation}\label{eq:coin-aux-sum-eq2-thtth-nonempty}
  S + S \,\mathtt{TTH} = M \,\mathtt{THTTH} + \, \mathtt{THTTH} \, ,
\end{equation}
where both sides are the sum of all sequences that either end with the first occurrence of the pattern, or end with the first two immediate reoccurrences of the pattern.
For sequences generated by independent Bernoulli trials, \autoref{eq:Bernoulli-aux-sum-M} and \autoref{eq:coin-aux-sum-eq2-thtth-nonempty} are the only two equations we need to solve for $S$.
Equivalently, we can split the auxiliary sum $M$ into $M_{\mathtt{H}}$ and $M_{\mathtt{T}}$ (cf., \autoref{fig:Bernoulli-GF-H-to-HH-and-T-to-HT}).
For sequences generated by first-order Markov trials, we split the auxiliary sum $M$ into $M_{\mathtt{A}}$ and $M_{\mathtt{R}}$, so that we can apply the memoryless property with the probability of alternation $p_A$ (cf., \autoref{fig:MarkovChains}).
In all of these cases, this type of event segmentation allows us to partition a pattern event into segments that are temporally independent.
It shows that with everything else fixed, the generating function $S$ is completely determined by pattern overlap.

\subsection{Pattern Overlap and Sample Length}
\autoref{eq:coin-aux-sum-eq2-thtth-nonempty} makes clear that a structural asymmetry between two random patterns can be captured by finite sample length, if the sample length allows overlapped reoccurrences of one pattern but not the other.
To illustrate, \autoref{tab:CMS-length3} lists two ways of counting patterns $\mathtt{HH}$ and $\mathtt{HT}$ within sequences of length $3$ generated by fair-coin Bernoulli trials.
Let $N(x)$ denote the number of occurrences for pattern $x$ within each sequence, we have $E[N(\mathtt{HH})] = E[N(\mathtt{HT})] = \sfrac{1}{2}$ over $8$ rows, but $N(\mathtt{HT})$ is more evenly distributed across the rows than $N(\mathtt{HH})$.
This is exactly the same result from \autoref{eq:Markov-EV-lump} or \autoref{eq:Bernoulli-EV-lump} that the inter-arrival times of $\mathtt{HH}$ and $\mathtt{HT}$ have the same mean but different variances.
Let $M(x)$ denote whether pattern $x$ occurs at least once and $\rho_x^{(n)}$ denote the probability of occurrence at least once within a sequence of length $n$, we have $\rho_{\mathtt{HH}}^{(3)} = E[M(\mathtt{HH})] = \sfrac{3}{8}$ and $\rho_{\mathtt{HT}}^{(3)} = E[M(\mathtt{HT})] = \sfrac{1}{2}$.
A comparison between $N(\mathtt{HH})$ and $M(\mathtt{HH})$ shows that $M(\mathtt{HH})$ discounts the overlapped reoccurrence of $\mathtt{HH}$ in sequence $\mathtt{HHH}$, which explains why $\rho_{\mathtt{HH}}^{(3)} < \rho_{\mathtt{HT}}^{(3)}$.
In all of these statistics, an asymmetry is revealed by the fact that $\mathtt{HH}$ can occur twice in a sequence of length $3$ but $\mathtt{HT}$ cannot.
In other words, we only need a sample length of $3$ to distinguish $\mathtt{HH}$ from $\mathtt{HT}$.

\begin{table}[htb]\centering\small
\caption{Two ways of counting patterns $\mathtt{HH}$ and $\mathtt{HT}$ within sequences of length $3$ generated by fair-coin Bernoulli trials.
$N(x)$ denotes the number of occurrences for pattern $x$, and $M(x) = 0 \text{ or } 1$ denotes whether pattern $x$ occurs at least once.
The alternation bias is revealed by different distributions of $N(\mathtt{HH})$ and $N(\mathtt{HT})$ across rows, and different averages between $M(\mathtt{HH})$ and $M(\mathtt{HT})$.
By both measures, an asymmetry can be observed within a sample length of $3$.
}
  \begin{tabular}{cccccc}
  \toprule
   Sequence & $N(\mathtt{HH})$ & $N(\mathtt{HT})$ & & $M(\mathtt{HH})$ & $M(\mathtt{HT})$  \\
  \midrule
   \texttt{TTT} & $0$ & $0$ & & $0$ & $0$ \\
   \texttt{TTH} & $0$ & $0$ & & $0$ & $0$ \\
   \texttt{THH} & $1$ & $0$ & & $1$ & $0$ \\
   \texttt{THT} & $0$ & $1$ & & $0$ & $1$ \\ \midrule
   \texttt{HTH} & $0$ & $1$ & & $0$ & $1$ \\
   \texttt{HTT} & $0$ & $1$ & & $0$ & $1$ \\
   \texttt{HHT} & $1$ & $1$ & & $1$ & $1$ \\
   \texttt{HHH} & $2$ & $0$ & & $1$ & $0$ \\ \midrule
   Average     & $\sfrac{1}{2}$ & $\sfrac{1}{2}$ & & $\sfrac{3}{8}$ & $\sfrac{1}{2}$ \\
  \bottomrule
  \end{tabular}
\label{tab:CMS-length3}
\end{table}

\subsection{Probabilities of First Occurrence and Occurrence At Least Once}
We now show that the result of a generating function can be approximated by recursively applying the overlap property within finite sample length.
We first define the \emph{probability of first occurrence}, $p_x^{(n)}$, as the probability that pattern $x$ arrives at the $n$th trial for the first time since the beginning of a counting process.
Let $T_{x|\varnothing}$ denote the time that pattern $x$ first arrives at $n$, we have,
\begin{equation}\label{def:p-first}
  p_x^{(n)} \equiv \mathrm{Pr}(T_{x|\varnothing} = n), \quad n \geq 1.
\end{equation}
Let $p_x$ denote the \emph{probability of occurrence} that pattern $x$ of length $k$ occurs in any $k$ consecutive flips.
By \autoref{eq:Markov-EV-lump} or \autoref{eq:Bernoulli-EV-lump}, $p_x$ is the inverse of the pattern's mean time.
For independent Bernoulli trials, we have $p^{(1)}_{\mathtt{HH}} = p^{(1)}_{\mathtt{HT}} = 0$, $p^{(2)}_{\mathtt{HH}} = p_{\mathtt{HH}} = p_{\mathtt{H}}^2$, and $p^{(2)}_{\mathtt{HT}} = p_{\mathtt{HT}} = p_{\mathtt{H}} p_{\mathtt{T}}$.
For pattern $\mathtt{HH}$ at $n > 2$,
\begin{equation}\label{eq:p-first-HH}
  p^{(n)}_{\mathtt{HH}} = p_{\mathtt{HH}} - p^{(n-1)}_{\mathtt{HH}} p_{\mathtt{H}}
  - \sum_{i=1}^{n-2} p^{(i)}_{\mathtt{HH}} \, p_{\mathtt{HH}}, \quad n > 2,
\end{equation}
where the term $p^{(n-1)}_{\mathtt{HH}} p_{\mathtt{H}}$ is the probability of overlapped reoccurrences when $\HH$ first arrives at $(n-1)$ then arrives again at $n$, and the term $\sum_{i=1}^{n-2} p^{(i)}_{\mathtt{HH}} \, p_{\mathtt{HH}}$ sums up all probabilities of non-overlapped reoccurrences.
Pattern $\mathtt{HT}$ has no overlapped reoccurrences, therefore
\begin{equation}\label{eq:p-first-HT}
  p^{(n)}_{\mathtt{HT}} = p_{\mathtt{HT}} - \sum_{i=1}^{n-2} p^{(i)}_{\mathtt{HT}} \, p_{\mathtt{HT}}, \quad n > 2.
\end{equation}

\begin{figure}[htb]\centering
\includegraphics[scale=1]{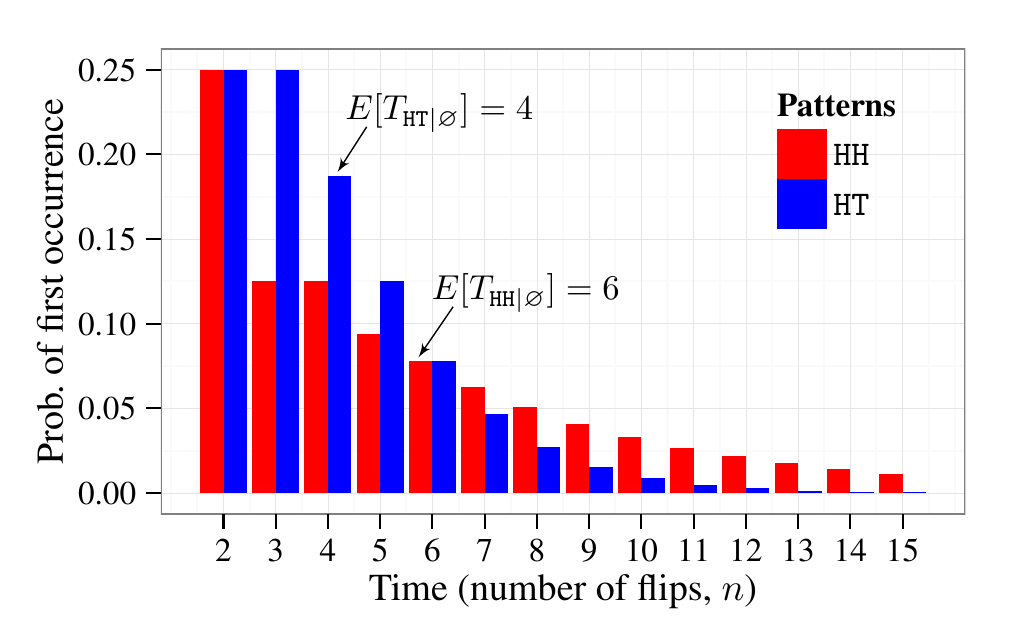}
\caption{Probabilities of first occurrence $p^{(n)}_{\mathtt{HH}}$ and $p^{(n)}_{\mathtt{HT}}$, by which the pattern first arrives at the $n$th flip in a sequence generated by fair-coin Bernoulli trials.}
\label{fig:CMS-P-first}
\end{figure}

\begin{figure}[htb]\centering
\includegraphics[scale=0.8]{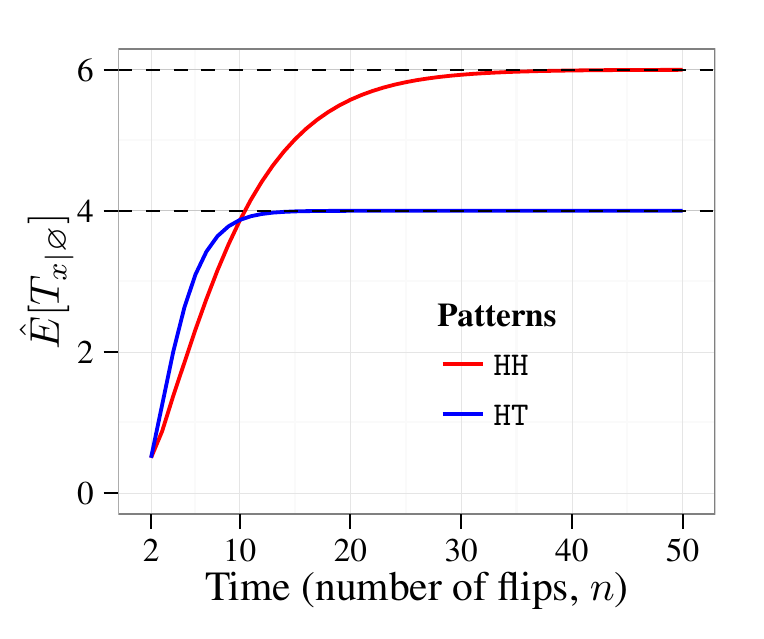}
\includegraphics[scale=0.8]{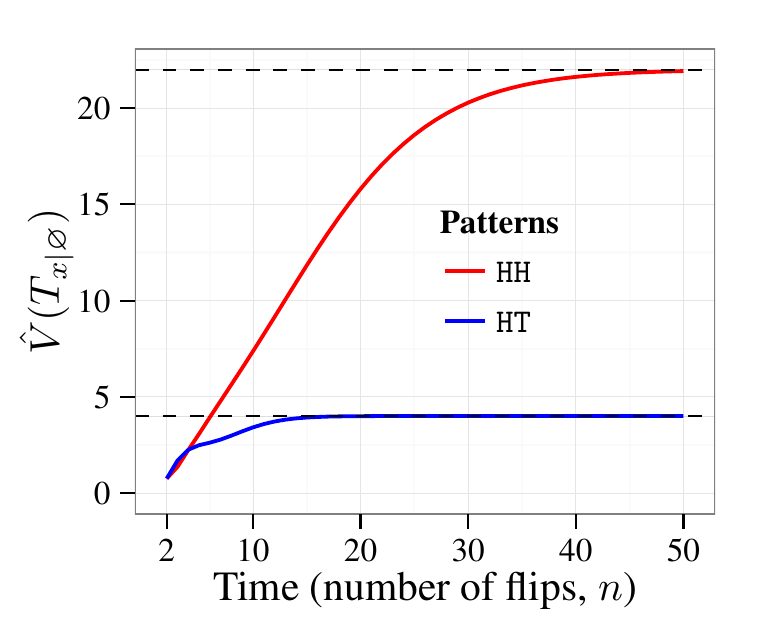}
\caption{Approximated expected values and variances of the first-arrival times for patterns $\mathtt{HH}$ and $\mathtt{HT}$ by fair-coin Bernoulli trials with finite sample length ({\protect \autoref{eq:wt-by-p-first}}).}
\label{fig:approx}
\end{figure}

\autoref{fig:CMS-P-first} shows the distribution of $p^{(n)}_{\mathtt{HH}}$ and $p^{(n)}_{\mathtt{HT}}$ over $n$ at $p_{\mathtt{H}} = \sfrac{1}{2}$.
Both $p^{(n)}_{\mathtt{HH}}$ and $p^{(n)}_{\mathtt{HT}}$ approach zero as $n$ increases, but $p^{(n)}_{\mathtt{HH}}$ drops more slowly, indicating that $T_{\mathtt{HH}|\varnothing}$ has a greater mean and a greater variance than $T_{\mathtt{HT}|\varnothing}$.
Indeed, for any pattern $x$, the mean and variance of its first arrival times can be approximated with the probability of first occurrence,
\begin{equation}\label{eq:wt-by-p-first}
  E[T_{x|\varnothing}] = \sum_{n=1}^{\infty} n p_x^{(n)},
  \quad \var(T_{x|\varnothing}) = \sum_{n=1}^{\infty} n^2 p_x^{(n)} - \big(\sum_{n=1}^{\infty} n p_x^{(n)}\big)^2.
\end{equation}
The closed forms of \autoref{eq:wt-by-p-first} for patterns $\mathtt{HH}$ and $\mathtt{HT}$ have been given by \autoref{eq:Bernoulli-EV-lump}, which shows that at $p_{\mathtt{H}}=\sfrac{1}{2}$, $E[T_{\mathtt{HH}|\varnothing}] = 6$, $\var(T_{\mathtt{HH}|\varnothing}) = 22$, $E[T_{\mathtt{HT}|\varnothing}] = 4$, and $\var(T_{\mathtt{HT}|\varnothing}) = 4$.
\autoref{fig:approx} shows that $n$ does not need to be very large to have a good approximation of \autoref{eq:wt-by-p-first}, for example, at $n=20$, $\hat{E}[T_{\mathtt{HH}|\varnothing}] = \sum_{n=1}^{20} n p_{\mathtt{HH}}^{(n)} \approx 5.574$, and $\hat{E}[T_{\mathtt{HT}|\varnothing}] = \sum_{n=1}^{20} n p_{\mathtt{HT}}^{(n)} \approx 3.999$.

\begin{figure}[htb]\centering
\includegraphics[scale=1]{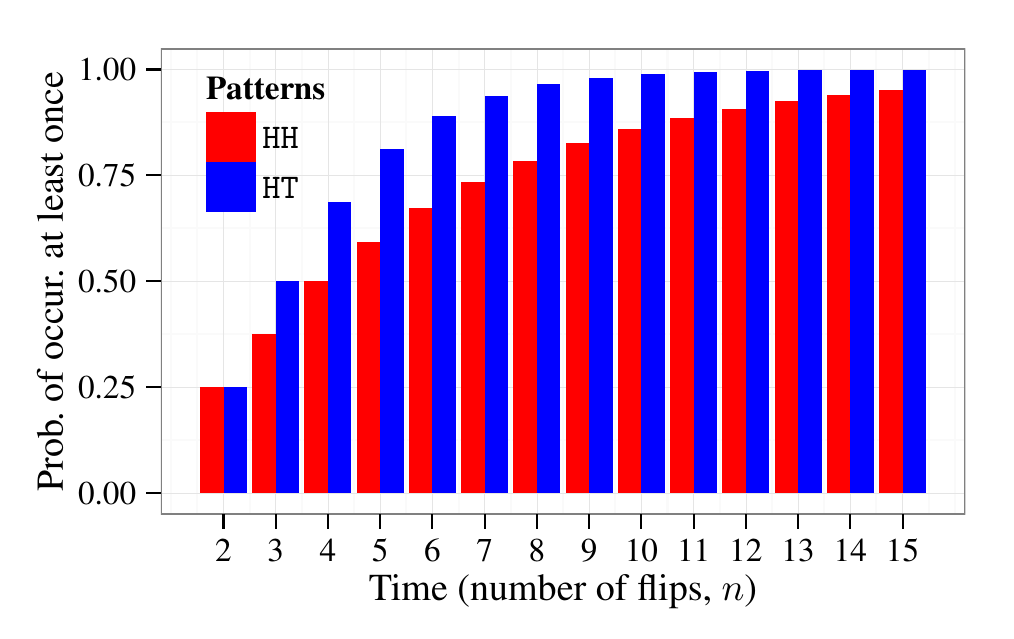}
\caption{Probabilities of occurrence at least once $\rho^{(n)}_{\mathtt{HH}}$ and $\rho^{(n)}_{\mathtt{HT}}$, by which a pattern occurs at least once within a sequence of length $n$ by fair-coin Bernoulli trials ({\protect \autoref{eq:p-at-least-once}}).}
\label{fig:CMS-P-GO}
\end{figure}

Because the probabilities of first occurrence are mutually exclusive at each $n$, we have the probability of occurrence at least once for pattern $x$ within a sequence of length $n$,
\begin{equation}\label{eq:p-at-least-once}
  \rho_x^{(n)} = \sum_{i=1}^n p_x^{(i)}.
\end{equation}\autoref{fig:CMS-P-GO} shows the distributions of $\rho^{(n)}_{\mathtt{HH}}$ and $\rho^{(n)}_{\mathtt{HT}}$ at $p_{\mathtt{H}} = \sfrac{1}{2}$, where both $\rho^{(n)}_{\mathtt{HH}}$ and $\rho^{(n)}_{\mathtt{HT}}$ approach probability one as $n$ increases, but $\rho^{(n)}_{\mathtt{HH}}$ does so more slowly than $\rho^{(n)}_{\mathtt{HT}}$ for the same reason explained by \autoref{eq:wt-by-p-first}.
At $n=3$, we have $\rho^{(3)}_{\mathtt{HH}} = \sfrac{3}{8}$, and $\rho^{(3)}_{\mathtt{HT}} = \sfrac{1}{2}$, which are respectively the expected values of $M(\mathtt{HH})$ and $M(\mathtt{HT})$ in \autoref{tab:CMS-length3}.

To recapitulate, both the probability of first occurrence and probability of occurrence at least once reveal a structural asymmetry between $\mathtt{HH}$ and $\mathtt{HT}$ with finite sample length.
To some extent, this indicates that limited memory capacity and limited exposure to random sequences may actually facilitate an early formation of the alternation bias in human perception of randomness.
As the sample length approaches infinity, the differences $p^{(n)}_{\mathtt{HH}} -p^{(n)}_{\mathtt{HT}}$ and $\rho^{(n)}_{\mathtt{HH}}-\rho^{(n)}_{\mathtt{HT}}$ approach zero (\autoref{fig:CMS-P-first} and \autoref{fig:CMS-P-GO}), but at the same time the difference $\hat{E}[T_{\mathtt{HH}|\varnothing}] - \hat{E}[T_{\mathtt{HT}|\varnothing}]$ approaches the maximal value of $2$ at $p_{\mathtt{H}}=\sfrac{1}{2}$ (\autoref{fig:approx}).

Furthermore, \autoref{tab:CMS-length3} and \autoref{eq:wt-by-p-first} show that the asymmetry between $\mathtt{HH}$ and $\mathtt{HT}$ can also be approximated by variances with finite sample length.
For example, the difference in the variances of inter-arrival times $\var(T_{\mathtt{HH|H}}) > \var(T_{\mathtt{HT|T}})$ means that reoccurrences of $\HH$ are more bunched together with greater spacing between, but reoccurrences of $\HT$ are more evenly distributed over time.
In behavioral economics, variance is often associated with risk \cite{Markowitz1952,Rabin2010,Sharpe1994,Weber2004risk}.
Then, the preference for an alternation pattern over a streak pattern may be interpreted as a consequence of risk aversion \cite{Sun2010jdm}.

\section{Conclusion}

In this paper, we show that the method of generating functions captures asymmetric temporal structures embedded in random sequences with multiple levels of abstraction.
Specifically, a generating function organize combinatorial objects with a simple ``juxtaposition'' arithmetic so that similarity-based structures can be converted to relational structures.
Then, a probability generating function compresses the relational structures into a time-invariant representation from which an abstract statistic can be extracted.
In addition, the results of generating functions are readily observable within finite sample length as they can be approximated by recursively applying overlapped representations.

Learning temporal structures via generating functions may shed new insights on some newly proposed learning mechanisms in cognitive neuroscience and AI.
In particular, it underscores the notions of distributed representations of random events across populations of neurons and multiple levels of abstraction by recurrent processing over time \cite{Elman1990,LeCun2015-NatureDeepReview,OReilly2014TI}.
It has been suggested that a powerful driving force behind the human intelligence is to learn by constantly predicting what will happen next and maximizing the compatibility between the internal representational state and the new inputs \cite{Hawkins2004}, and such predictive learning may be implemented by a neural algorithm of temporal integration, supported by the deep versus superficial layers of neocortex and their interconnections with the thalamus \cite{OReilly2014TI}.
Indeed, we have shown that with unsupervised learning, a biologically plausible neural network is capable of learning meaningful temporal structures such as $\atime{HH}{H} > \atime{HT}{H}$ and $\var(T_{\mathtt{HH|H}}) > \var(T_{\mathtt{HT|T}})$ \cite{Sun2015pnas}.

Finally, the method of generating functions not only produces pattern time statistics, but also gives a coherent interpretation to other statistical measures such as probabilities of occurrence, first occurrence and occurrence at least once.
As illustrated by the Markov chains in \autoref{fig:MarkovChains} and \autoref{fig:Bernoulli-GF-H-to-HH-and-T-to-HT}, different measures may arise due to different initial conditions and different ways of event segmentation.
In this regard, this method corresponds well with the rich statistical representations in the human brain, the effectiveness of predictive learning and the sensitivity of the human mind to the latent structures in the learning environment.

\phantomsection


\end{document}